\documentclass[letterpaper]{jpconf}
\usepackage{graphicx}

\begin{document}

\title{The OPERA experiment:~Preliminary results from the 2008 run}

\author{Guillaume Lutter on behalf of the OPERA Collaboration.}

\address{Centre for Research and Education in Fundamental Physics, Laboratory for High Energy Physics (LHEP), University of Berne.}

\begin{abstract}
The neutrino flavour oscillations hypothesis has been confirmed by several experiments, all are based on the observation of the disappearance of a given neutrino flavour. The long baseline neutrino experiment OPERA (Oscillation Project with Emulsion tRacking Apparatus) aims to give the first direct proof of the $\tau$ neutrino appearance in a pure muon neutrino beam (CERN Neutrinos to Gran Sasso beam). In 2008 the OPERA experiment has started full data taking with the CNGS beam and around 1700 interactions have been recorded. The experiment status and the first results from the 2008 run are presented.
\end{abstract}

\section{Introduction}
The OPERA experiment, located in the Gran Sasso underground laboratory (LNGS) in Italy, is a long-baseline experiment designed to obtain an unambiguous signature of $\nu_{\mu} \rightarrow \nu_{\tau}$ oscillations in the parameter region indicated by the atmospheric neutrino experiments \cite{proposal}. 
The detector developed by the international collaboration OPERA is designed to search primary for $\nu_{\tau}$ appearance in the high energy $\nu_{\mu}$ CERN to Gran Sasso (CNGS) beam at 730 km from the neutrino source. It may also explore the $\nu_{\mu} \rightarrow \nu_e$ oscillation channel and improve the limits on the third yet unknown mixing angle $\theta_{13}$. 

The $\nu_{\tau}$ direct appearance search is based on the observation of charge current interaction (CC) events with the $\tau$ decaying through leptonic and hadronic channels. The principle of the OPERA experiment is to observe the $\tau$ trajectories and the decay products. Because the weak neutrino cross section and the short $\tau$ lifetime, the OPERA detector must combine a huge mass with a high granularity which can be achieved by using nuclear emulsions.

\section{The OPERA detector}

The OPERA detector is composed of two identical parts called Super Module (SM) \cite{techpaper}. Each SM has a target section and a muon spectrometer [Fig. 1].

\begin{figure}[hbtc]
	\centering
	\label{figdetec}
		\includegraphics[width=.8\textwidth]{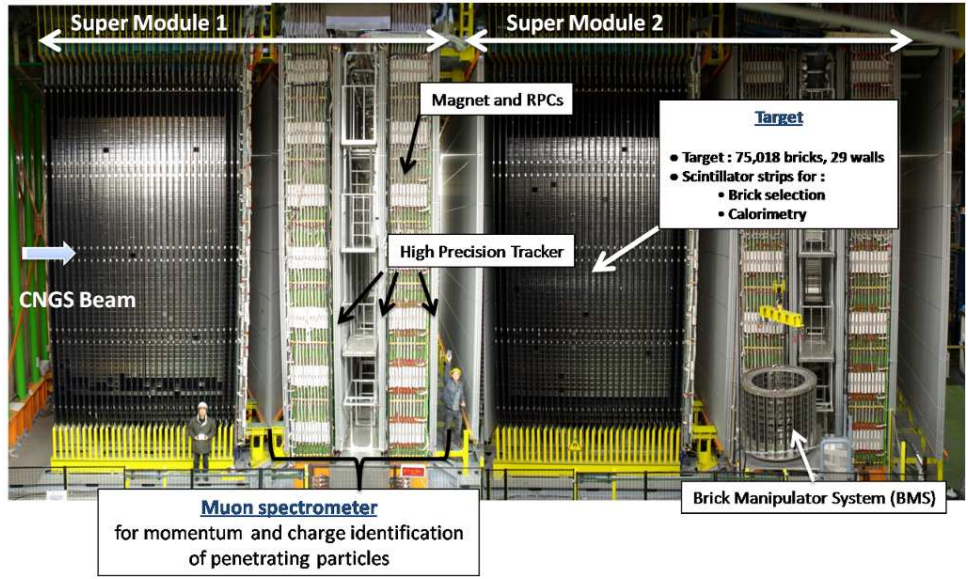}
		\caption{View of the OPERA detector.}
\end{figure}

The target section is composed of 29 vertical supporting steel structures called walls. The walls contain the basic target detector units called Emulsion Cloud Chambers (ECC) brick. The total OPERA target contains 150036 ECC bricks with a total mass of 1.25 ktons. Each ECC brick is a sequence of 57 emulsion films interleaved with 56 lead plates (1 mm thick). An emulsion film is composed of a pair of 44 $\mu m$ thick emulsion layers deposited on a 205 $\mu m$ plastic base. The ECC bricks have been assembled underground at an average rate of 700 per day by a dedicated fully automated Brick Assembly Machine (BAM) and the OPERA target has been filled by using two automated manipulator systems (BMS).

Downstream of each brick [Fig. 2], an emulsion film doublet called Changeable Sheet (CS) is attached in a separate enveloppe. The CS doublet can be detached from the brick for analysis to confirm and locate the tracks produced in the electronic detectors by neutrino interactions. The CS doublet is the interface between the ECC brick and the electronic detector. Indeed, each wall is interleaved with a double layered wall of scintillator strips. This electronic detector called Target Tracker (TT) also provides a trigger for the neutrino interactions.

The spectrometer allows the determination of the muons charge and momentum by measuring their curvature in a dipolar magnet made of iron. Each spectrometer is equipped with RPC bakelite chambers and High Precision Tracker (HPT) composed of drift-tubes. The spectrometer reduces the charge confusion to less than 0.3 \%, gives a muon momentum measurement better than 20\% for a momentum less than 50 GeV and reaches a muon identification efficiency of 95 \%. 

\begin{figure}[hbtc]
	\centering
	\label{figbrickcs}
		\includegraphics[width=.25\textwidth]{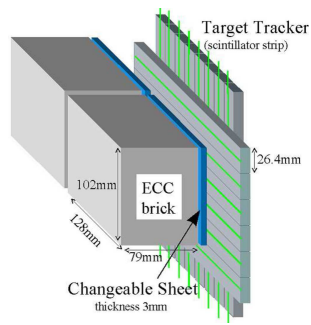}
		\caption{Schematic view of two bricks with their  Changeable Sheets and target tracker planes.}
\end{figure}

\section{The CNGS beam and projected results}

The CERN to Gran Sasso (CNGS) $\nu_{\mu}$ beam is designed and optimized to maximize the number of charged current interactions of $\nu_{\tau}$ produced by oscillation at LNGS. With $4.5 \times 10^{19}$ protons on target per year, the number of CC and neutral current (NC) interactions expected in the Gran Sasso laboratory from $\nu_\mu$ are respectively about 2900 per kton per year and 875 per kton per year. If the $\nu_\mu \rightarrow \nu_\tau$ oscillation hypothesis is confirmed, the number of $\tau$'s observed in the OPERA detector after 5 years of data taking is expected to be 10 events with a background of 0.75 events for a $\Delta m^2= 2.5 \times 10^{-3}eV^2$ at full mixing. The OPERA detector events are synchronized with the CNGS beam using a sophisticated GPS system.

\section{The OPERA strategy}

For a given event, the electronic detectors give the map of probable locations for the interaction brick. The brick with the highest probability is extracted by the BMS. Then the CS doublet is detached from the brick and developed in the underground facility. The two emulsion films are scanned by using fast automated microscopes. If the track candidates found on the CS doublet match with the electronic data then the brick is exposed for 12 hours to cosmic rays in order to help with the film-to-film alignment in the brick. Subsequently the brick is developed in an automated facility and sent to the scanning laboratories either in Europe or in Japan.
The brick scanning is done by computer driven fast microscopes. The vertex finding strategy consists in following back, film by film, the tracks found on the CS doublet untill the tracks stop inside the brick. The scanning speed can reach up to 20 $cm^2/h$ while keeping a good spatial and angular resolution. To confirm the stopping track, an area scan of several $mm^2$  around the stopping point of the tracks is performed for 5 films upstream and downstream. Then a vertex interaction can be reconstructed and a topology compatible with the decay of a $\tau$ lepton is searched for.

\section{Status of the 2008 run}

After a short commissioning run in 2006, the CNGS operation started on September 2007 at rather low intensity with 40\% of the total OPERA target mass. Since the CNGS encountered operational problems the physics run lasted only a few days. During this run $0.082 \times 10^{19}$ protons on target (p.o.t.) were accumulated and 465 events were recorded, of which 35 in the target region.

From June to November 2008, $1.782 \times 10^{19}$ p.o.t. were delivered by the CNGS \cite{2008run}. OPERA collected 10100 events and among them 1663 interactions in the target region where 1723 were expected. The other events originated in the spectrometers, the supporting structures, the rock surrounding the cavern, the hall structure.
All electronic detectors were operational and the live time of the data acquisition system exceeded 99\%.

For the events classified as CC interations in the target [Fig. 3] the muon momentum and the muon angle in the vertical (y-x) plane with respect to the horizontal (z) axis distributions are compared to the Monte Carlo (MC) expectation.
The beam direction angle is found to be tilted by 58 mrad as expected from the geodesy.

\begin{figure}[hbtc]
	\centering
	\label{fig:distri}
		\includegraphics[width=.7\textwidth]{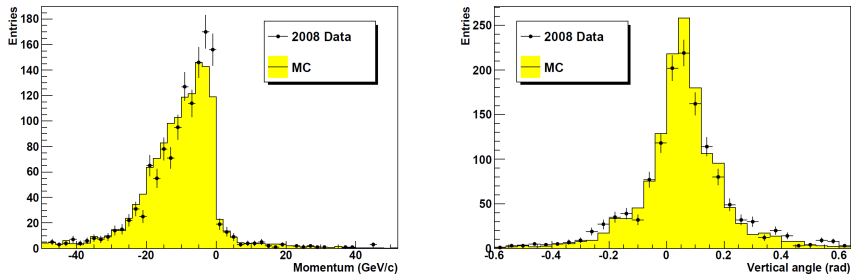}
		\caption{Left: momentum distribution of muons produced in CC neutrino interactions inside OPERA target. Right: angular distribution of the muon tracks with respect to the horizontal axis.}
\end{figure}

In the beginning of April 2009, 1038 bricks have been developed and around 700 events have been located. The brick finding effiency is $88.3 \pm 5 \%$ and the vertex finding efficiency in the selected bricks for CC events is between 84-95 \% with 93\% predicted by MC and for NC events between 70-91 \% with 81\% expected from MC.

Among the located events, 7 events present a charm-like decay topology in agreement with 9.29 events as predicted by the Monte Carlo simulations. Charm production and decay topology events have a great importance in OPERA. Indeed, the charm decays exhibit the same topology as $\tau$ decays and they are a potential source of background if the muon at the primary vertex is not identified. 
A charm-like topology is shown in figure [Fig. 4] where a track presents a decay-kink.

\begin{figure}[hbtc]
	\centering
	\label{fig:charm}
		\includegraphics[width=.7\textwidth]{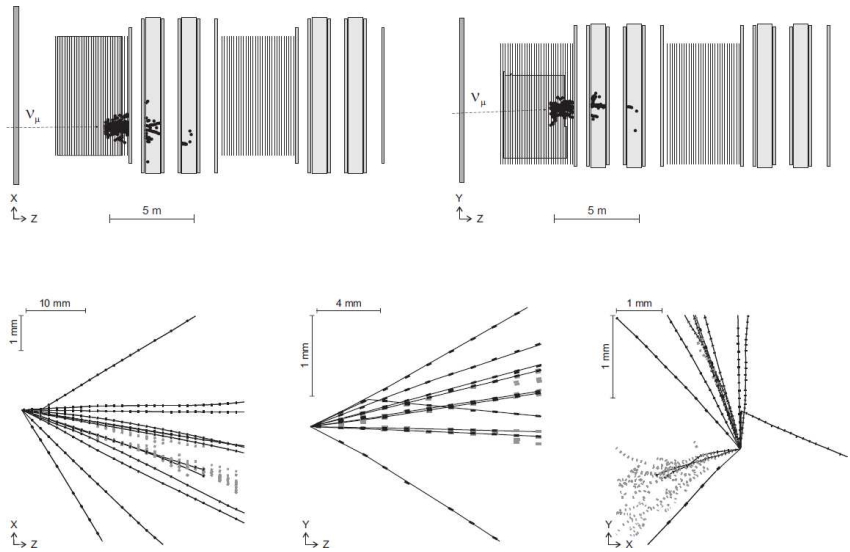}
		\caption{Display of the OPERA electronic detector of a $\nu_\mu$ charged-current interaction with a charm like topology (top panel). The emulsion reconstruction is shown in the bottom panels where the charm-like topology is seen with a kink: top view (bottom left), side view (bottom center), frontal view (bottom right). The dots in the lower panel are due to an electromagnetic shower.}
\end{figure}

\section{Conclusion}

During the 2008 CNGS run all the electronic detectors performed well. The OPERA strategy has been validated and the vertex location was successfully accomplished for CC and NC events. In the analysed data sample, 7 events with a charm-like topology were found. This is consistent with the expectation and shows the success of combining the topological and kinematical analyses.
The 2008 run constitute an important milestone for the OPERA experiment. For the 2009 run, around $3.5 \times 10^{19}$ p.o.t. are expected. The integrated statistics would be sufficient to expect the observation of two $\tau$ events and give a precise estimation of detector efficiency, background and sensitivity.

\end{document}